\documentclass[twocolumn]{aastex6}
\pdfoutput=1 
\usepackage{amsmath,amstext}
\usepackage[T1]{fontenc}
\usepackage{apjfonts} 
\usepackage[figure,figure*]{hypcap}
\usepackage{microtype}

\usepackage{xcolor}

\shorttitle{Radio Emission from GW\,170817}
\shortauthors{Alexander et al.}

\begin{document}

\title{The Electromagnetic Counterpart of the Binary Neutron Star Merger LIGO/VIRGO GW170817.\\ VI. Radio Constraints on a Relativistic Jet and Predictions for Late-Time Emission from the Kilonova Ejecta}

\author{K.~D.~Alexander\altaffilmark{1}, E.~Berger\altaffilmark{1}, W.~Fong\altaffilmark{2,3}, {P.~K.~G.~Williams}\altaffilmark{1}, {C.~Guidorzi}\altaffilmark{4},{R.~Margutti}\altaffilmark{3}, {B.~D.~Metzger}\altaffilmark{5}, {J.~Annis}\altaffilmark{6}, {P.~K.~Blanchard}\altaffilmark{1}, {D.~Brout}\altaffilmark{7}, {D.~A.~Brown}\altaffilmark{8}, {H.-Y.~Chen}\altaffilmark{9}, {R.~Chornock}\altaffilmark{10}, {P.~S.~Cowperthwaite}\altaffilmark{1}, {M.~Drout}\altaffilmark{2,11},{T.~Eftekhari}\altaffilmark{1}, {J.~Frieman}\altaffilmark{6,9},{D.~E.~Holz}\altaffilmark{9,12}, {M.~Nicholl}\altaffilmark{1}, {A.~Rest}\altaffilmark{13,14}, {M.~Sako}\altaffilmark{7}, {M.~Soares-Santos}\altaffilmark{6}, {V.~A.~Villar}\altaffilmark{1}}

\altaffiltext{1}{Harvard-Smithsonian Center for Astrophysics, 60 Garden Street, Cambridge, Massachusetts 02138, USA}
\altaffiltext{2}{Hubble Fellow}
\altaffiltext{3}{Center for Interdisciplinary Exploration and Research in Astrophysics (CIERA) and Department of Physics and Astronomy, Northwestern University, Evanston, IL 60208}
\altaffiltext{4}{Department of Physics and Earth Science, University of Ferrara, via Saragat 1, I--44122, Ferrara, Italy}
\altaffiltext{5}{Department of Physics and Columbia Astrophysics Laboratory, Columbia University, New York, NY 10027, USA}
\altaffiltext{6}{Fermi National Accelerator Laboratory, P.O. Box 500, Batavia, IL 60510, USA}
\altaffiltext{7}{Department of Physics and Astronomy, University of Pennsylvania, Philadelphia, PA 19104, USA}
\altaffiltext{8}{Department of Physics, Syracuse University, Syracuse NY 13224, USA}
\altaffiltext{9}{Kavli Institute for Cosmological Physics, University of Chicago, Chicago, IL 60637, USA}
\altaffiltext{10}{Astrophysical Institute, Department of Physics and Astronomy, 251B Clippinger Lab, Ohio University, Athens, OH 45701, USA}
\altaffiltext{11}{Carnegie Observatories, 813 Santa Barbara Street, Pasadena, CA 91101, USA}
\altaffiltext{12}{Enrico Fermi Institute, Department of Physics, Department of Astronomy and Astrophysics, 5640 South Ellis Avenue, Chicago, IL 60637 USA}
\altaffiltext{13}{Space Telescope Science Institute, 3700 San Martin Drive, Baltimore, MD 21218, USA}
\altaffiltext{14}{Department of Physics and Astronomy, The Johns Hopkins University, 3400 North Charles Street, Baltimore, MD 21218, USA}

\begin{abstract}
We present Very Large Array (VLA) and Atacama Large Millimeter/sub-millimeter Array ALMA radio observations of GW\,170817, the first Laser Interferometer Gravitational-wave Observatory (LIGO)/Virgo gravitational wave (GW) event from a binary neutron star merger and the first GW event with an electromagnetic (EM) counterpart.  Our data include the first observations following the discovery of the optical transient at both the centimeter ($13.7$ hours post merger) and millimeter ($2.41$ days post merger) bands. We detect faint emission at 6 GHz at 19.47 and 39.23 days after the merger, but not in an earlier observation at 2.46 d. We do not detect cm/mm emission at the position of the optical counterpart at frequencies of 10-97.5 GHz at times ranging from 0.6 to 30 days post merger, ruling out an on-axis short gamma-ray burst (SGRB) for energies $\gtrsim 10^{48}$ erg. For fiducial SGRB parameters, our limits require an observer viewer angle of $\gtrsim 20^{\circ}$. The radio and X-ray data can be jointly explained as the afterglow emission from an SGRB with a jet energy of $\sim 10^{49}-10^{50}$ erg that exploded in a uniform density environment with $n\sim 10^{-4}-10^{-2}$ cm$^{-3}$, viewed at an angle of $\sim 20^{\circ}-40^{\circ}$ from the jet axis. Using the results of our light curve and spectral modeling, in conjunction with the inference of the circumbinary density, we predict the emergence of late-time radio emission from the deceleration of the kilonova (KN) ejecta on a timescale of $\sim 5-10$ years that will remain detectable for decades with next-generation radio facilities, making GW\,170817 a compelling target for long-term radio monitoring.
\end{abstract}

\keywords{gravitational waves ---  
relativistic processes}

\section{Introduction}

Radio emission from binary neutron star (BNS) mergers offers a unique way to probe the energetics and geometries of their outflows, as well as their circumbinary densities.  Until now, the primary way to place constraints on radio emission from BNS mergers was through rapid follow-up observations of short-duration gamma-ray bursts (SGRBs), which have been argued to result from BNS mergers (e.g., \citealt{ber14}). A decade of SGRB radio follow-up at GHz frequencies yielded four detections and multiple upper limits \citep{bpc+05,sbk+06,fbm+14,fbm+15}, providing tight constraints on the energetics and densities of the burst environments. In particular, these observations point to low-density environments of $\sim 10^{-3}-0.1$~cm$^{-3}$, and typical beaming-corrected energies of $\sim 10^{49}-10^{50}$ erg \citep{ber14,fbm+15}. 

The jets launched by SGRBs are collimated and highly relativistic, meaning that for their typical cosmological distances \citep{ber14} they are only detectable within a narrow range of viewing angles at early times \citep{fbm+15}. For nearby BNS mergers within 200~Mpc as expected from gravitational wave (GW) detections \citep{LVCdistance}, the emission could be detectable off-axis \citep{np11,mb12}.

On 2017 August 17 12:41:04 UTC, the Advanced Laser Interferometer Gravitational-wave Observatory (LIGO)/Virgo detected a gravitational wave signal determined to originate from a BNS at a distance of $\sim 40$ Mpc \citep{ALVgcn,ALVdetection}. The localization of GW\,170817 was spatially coincident with a weak gamma-ray transient detected by {\it Fermi}/GBM 
\citep{GBMgcn1,GBMgcn2,GBMgcn3,GBMdetection} and INTEGRAL \citep{INTEGRALgcn,INTEGRALdetection}, termed GRB\,170817A. Subsequently, an optical counterpart\footnote{This source is variously known as AT2017gfo (International Astronomical Union name), SSS17a \citep{SWOPEgcn,SWOPEpaper}, and DLT17ck \citep{DLT40gcn,DLT40Paper}.} was also discovered by several teams, including by our group with DECam \citep{SWOPEgcn,DECAMgcn,DLT40gcn,SWOPEpaper,DECamPaper1,DLT40Paper}.  These detections and localization make GW170817 the first GW event with an EM counterpart, ushering in the era of multi-messenger GW-EM astronomy \citep{GWEMcapstone}.

Here, we present centimeter and millimeter wavelength follow-up observations of the optical counterpart, starting $\approx 0.6$ days post-merger and extending to $\approx 39$ days. Together with our detailed X-ray analysis \citep{DECamPaper5}, we use the radio data to place tight constraints on the presence of an on- or off-axis jet. Finally, we present expectations for long-term radio emission produced by the ejecta powering the optical/NIR kilonova emission \citep{DECamPaper2,DECamPaper3,DECamPaper4}.  Our radio observations of the host galaxy, NGC\,4993, are discussed in \cite{DECamPaper8}.

\section{Observations}

We initiated radio observations of the position of the optical counterpart with the Karl G. Jansky Very Large Array (NRAO program VLA/17A-218; PI:~Fong) at 9.7 GHz beginning on 2017 August 18.10 UT (13.7 hr post merger). Subsequently, deeper VLA observations were obtained under the shared public program TTRA0001 (PI:~Mioduszewski) at 1.4 d (10 GHz) and at 2.4 d (6 GHz, 10 GHz, and 15 GHz).  All subsequent VLA observations, beginning at 5.5 d, were collected under program 17A-231 (PI:~Alexander). The observations were performed in C, C$\rightarrow$B, or B configurations. We analyzed and imaged the VLA data using standard CASA routines \citep{the.casa}, using 3C286 as the flux calibrator and J$1258{-}2219$ as the phase calibrator. For some epochs, we also compared our reduction to the calibration performed by the automated VLA pipeline as a cross-check. For all detected sources, we fit the flux density and position of the emission using the \textsf{imtool} program within the \textsf{pwkit} package\footnote{Available at \url{https://github.com/pkgw/pwkit}.}. The reported uncertainty for detections is the uncertainty on this fit, not the image RMS at the source position, which is reported separately by \textsf{imtool}. We use the fit uncertainty to estimate the significance of detection. The observations are summarized in Table \ref{tab:obs}.

At the position of the optical counterpart we do not detect emission with a signal-to-noise ratio of $\gtrsim 3\sigma$ in any of our observations at $< 19$ d. On 2017 September 2 and 3, \cite{mool} and \cite{corsi} reported the emergence of radio emission with the VLA at a signal-to-noise ratio of $\sim 5$ (summarized in \citealt{hal17}), which was tentatively confirmed by the Australia Telescope Compact Array (ATCA) \citep{ATCAgcn}. Our subsequent observations of similar duration on 2017 September 5 UT were affected by marginal weather conditions and our initial reduction showed only a very weak peak ($\sim2\sigma)$ that did not meet our standards for detection \citep{alex17}. We were subsequently able to improve the noise properties of our image and we detect marginal emission at 6 GHz with a flux density of $19\pm 6$ $\mu$Jy ($3.1\sigma$ significance). Given the weather impact, we conclude that this emission is likely consistent with the earlier detections reported by \cite{mool} and \cite{corsi}, despite the lower significance. The source is not detected to a comparable depth in our contemporaneous observations at 10 GHz, although bad weather disproportionately affects high-frequency observations so these data may suffer from flux decorrelation, as also suggested by the lower flux found for the host galaxy in this epoch. We combined our 10 GHz data taken on August 30 and September 5 and do not detect any radio emission to a $3\sigma$ limit of 11 $\mu$Jy. These observations suggest that the source spectral energy distribution is optically thin at this time. We continue to detect the source in observations at 6 GHz taken on 2017 September 25 UT, with a flux density of $27\pm6$ $\mu$Jy.

We also observed the position of the optical counterpart with the Atacama Large Millimeter/submillimeter Array (ALMA) beginning 2.4 d post merger (programs 2016.A.00043.T and 2016.A.00046.T; PI: Alexander). Additional observations were obtained at 9.4, 15.3, and 30.3~d.  The first two observations each lasted 30 min, while the second two lasted 1 hr each. In all cases, we used the Band 3 receiver system in wideband continuum mode, with two spectral windows of 4 GHz width centered at  frequencies of 91.5 and 103.5 GHz. We calibrated and imaged the data using a custom pipeline based on CASA \citep{the.casa}, using the quasar B$1334-127$ as the bandpass and flux density calibrator. The observations are summarized in Table~\ref{tab:obs}. We combined the data in both subbands and all four epochs and we do not detect emission at the optical transient position, with an image RMS of 8.5~$\mu$Jy at that position.

In all of our radio observations, we detect emission coincident with the optical center of the host galaxy, NGC\,4993. The host emission is unresolved in the ALMA data, with a beam size of 0.21" (corresponding to $\lesssim 40$ pc at a distance of 39.5 Mpc) and also appears unresolved in our VLA observations. There is no evidence for extended host radio emission at the position of the optical counterpart. For further discussion of the host emission we refer the reader to the companion paper, \citet{DECamPaper8}.

\begin{longtable*}{lcrrccccc}
\caption{Radio Observations of GW170817 and Its Host Galaxy, NGC 4993} 
\label{tab:obs} \\
\hline
Observatory & Start Date & $\Delta t$ & Avg Freq & {  Freq Range} & {  On-source} & {  Beam Size} & Img RMS & Host flux \\        
   			 & (UT)  & (d) & (GHz)   &  {  (GHz)} & {  Time (hr)} & {  (arcsec)} & ($\mu$Jy/beam)$^{a}$ &  density ($\mu$Jy) \\        
\hline
VLA & {  2017 Aug 18.1} & 0.57 & 9.7  & {  8.0--9.0, 10.5--11.5} & {  0.07} & {  9.0 $\times$ 1.5} & 48 & $250 \pm 55$\\
VLA & {  2017 Aug 18.9} & 1.44 & 10.0  & {  8.0--11.9} & {  1.5} & {  3.1 $\times$ 1.4}&  4.6 &$372 \pm 17$\\
ALMA & {  2017 Aug 19.9} & 2.41 & 97.5  & {  89.5--93.5, 101.5--105.5} & { 0.14} & {  0.3 $\times$ 0.2} & 25 & $174\pm 34$\\
VLA & {  2017 Aug 19.9} & 2.42 & 15.0  & {  12.0--17.9} & {  0.46} & {  2.2 $\times$ 1.0} & 5.9 &$295\pm 18$\\ 
VLA & {  2017 Aug 19.9} & 2.44 & 10.0  & {  8.0--11.9} & {  0.41} & {  3.1 $\times$ 1.4}&  5.7& $330\pm 11$\\ 
VLA & {  2017 Aug 19.9} & 2.46 & 6.0  & {  4.0--7.9} & {  0.41} & {  5.5 $\times$ 2.4}& 7.3 & $354\pm 12$\\
VLA & {  2017 Aug 23.0} & 5.48 & 10.0  & {  7.9--11.9} & {  0.60} & {  3.8 $\times$ 1.5} & 9.5& $274\pm22$\\ 
VLA & {  2017 Aug 25.8} & 8.29 & 10.0  & {  8.0--11.9} & {  0.62} &{  4.0 $\times$ 1.5} & 5.8 &$260\pm10$\\  
ALMA & {  2017 Aug 27.0} & 9.43 & 97.5  & {  89.5--93.5, 101.5--105.5} & { 0.13} & {  0.4 $\times$ 0.2} & 24 & $234\pm32$\\
VLA & {  2017 Aug 30.9} & 13.41 & 10.0  & {  8.0--11.9} & {  0.60} & {  1.0 $\times$ 0.5} & 6.1 & $262\pm20$\\  
ALMA & {  2017 Sep 1.8} & 15.33 & 97.5  & {  89.5--93.5, 101.5--105.5} & { 0.68} & {  0.2 $\times$ 0.2} & 13 & $253\pm16$\\
VLA & {  2017 Sep 5.9} & 19.43 & 10.0  & {  8.0--11.9} & {  0.93} & {  1.1 $\times$ 0.5} & 4.5 & $218\pm11$\\ 
VLA & {  2017 Sep 5.9} & 19.47 & 6.0  & {  4.0--7.9} & {  0.93} & {  2.2 $\times$ 0.8} & 4.0$^{b}$ & $345\pm12$\\  
ALMA & {  2017 Sep 16.9} & 30.34 & 97.5  & {  89.5--93.5, 101.5--105.5} & { 0.68} & {  0.2 $\times$ 0.1} & 14 & $223\pm23$\\
VLA & {  2017 Sep 25.7} & 39.23 & 6.0  & {  4.0--7.9} & {  1.9 }&{  1.7 $\times$ 0.7} & {  4.4}$^{c}$ & $314\pm11$ \\ 
\hline
\noalign{\smallskip}
\noalign{\text{$^{a}$At the position of the optical and X-ray counterpart. {  $3\sigma$ upper limits plotted in figures are 3 $\times$ RMS.}}}
\noalign{\text{$^{b}$Emission detected at the counterpart position with a flux density of $19\pm6$ $\mu$Jy.}}
\noalign{\text{$^{c}$Emission detected at the counterpart position with a flux density of $27\pm6$ $\mu$Jy.}}
\end{longtable*}

\section{Afterglow constraints}

The EM observations indicate that the BNS merger that produced GW170817 was accompanied by gamma-ray emission \citep{GBMgcn1,GBMgcn2,GBMgcn3,GBMdetection,INTEGRALgcn,INTEGRALdetection}. The shock wave produced between a SGRB jet and the surrounding medium generates a broadband synchrotron afterglow, which is expected to be the dominant source of radio emission within the first few months after the merger \citep{spn98,np11}. Here we utilize the standard afterglow synchrotron model in a constant density medium \citep{gs02}, as expected for a BNS progenitor. This model provides a mapping from the afterglow spectral energy distribution and temporal evolution to the isotropic-equivalent kinetic energy ($E_{\rm K,iso}$), circumbinary density ($n$), fractions of post-shock energy in radiating electrons ($\epsilon_e$) and magnetic fields ($\epsilon_B$), and the electron power-law distribution index ($p$), with $N(\gamma)\propto \gamma^{-p}$ for $\gamma\gtrsim \gamma_{\rm min}$, where $\gamma_{\rm min}$ is the minimum Lorentz factor of the electron distribution accelerated by the shock.

Since the jet is likely initially highly relativistic and collimated with a jet opening angle $\theta_j$
, the observed afterglow depends on the viewing angle of the observer with respect to the jet axis, $\theta_{\rm obs}$ \citep{gpk+02}. When $\theta_{\rm obs}\gtrsim \theta_{j}$, the emission is initially relativistically beamed away from the observer, and only becomes visible as the jet spreads and decelerates (e.g., \citealt{vm12}). The observed emission therefore peaks with a viewing angle-dependent delay of days to months, when compared to the on-axis case. Below, we consider separately the constraints our observations place on the SGRB properties in both on- and off-axis models.

\subsection{On-axis Afterglow Models}

\begin{figure} 
\centerline{\includegraphics[width=3.5in]{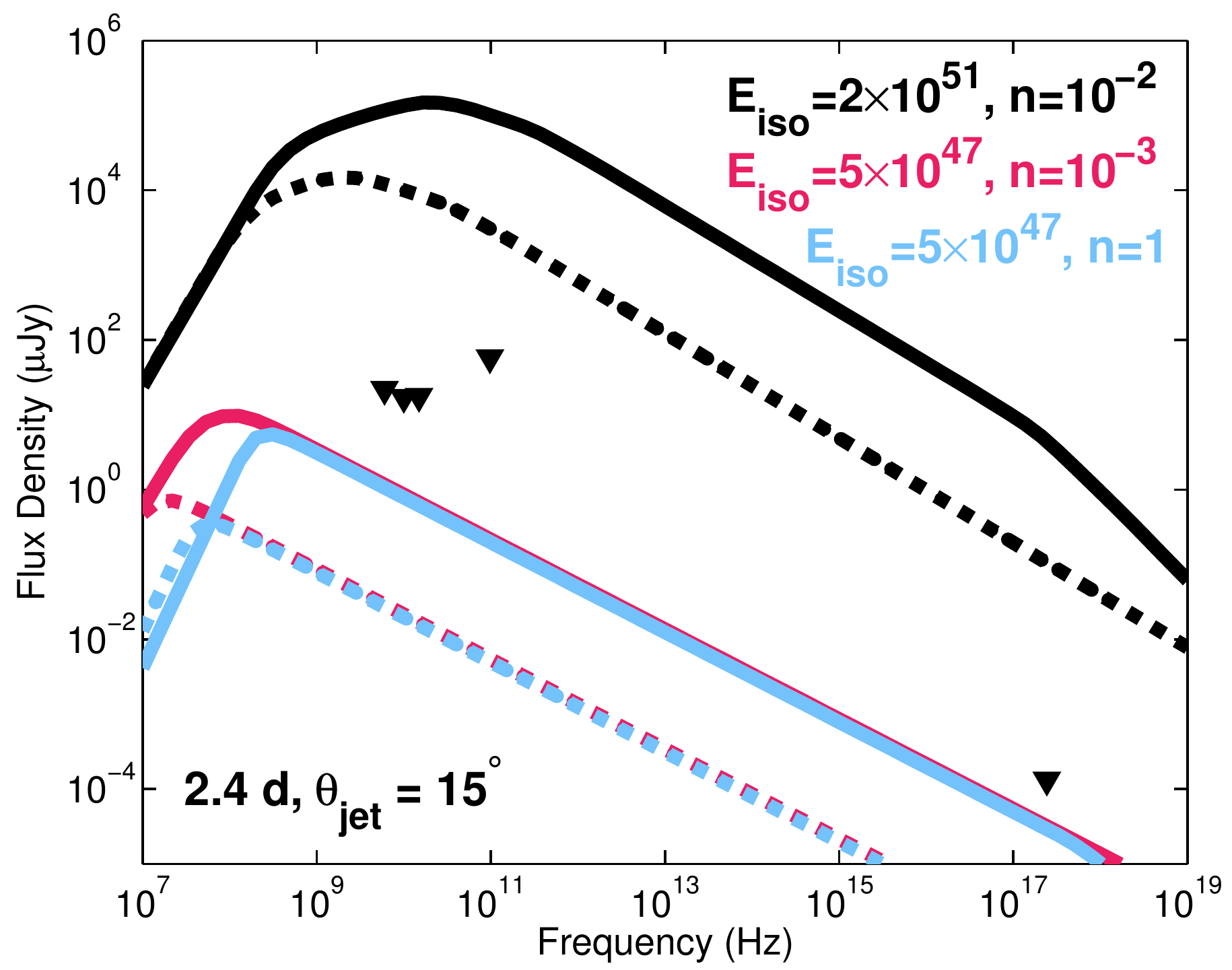}}
\caption{On-axis SGRB afterglow model spectral energy distributions at 2.4 d shown with our radio and X-ray upper limits at this epoch from the VLA, ALMA, and \textit{Chandra} (black triangles). The red and blue curves show allowed models with $E_{\rm K,iso}=5\times10^{47}$ erg for $n=10^{-3}$ cm$^{-3}$ and $n=1$ cm$^{-3}$, respectively. The predicted emission from a ``canonical'' on-axis SGRB (black) is shown for comparison. We show both $\epsilon_B=0.01$ (solid lines) and $\epsilon_B=10^{-4}$ (dashed lines). The flux density for fixed $\epsilon_B$ is nearly density-independent at radio and X-ray frequencies for the low-energy models, as the jet has already decelerated by the time of these observations.}
\label{fig:onaxis}
\end{figure}

We first consider whether our radio observations are consistent with the emission expected from an on-axis SGRB afterglow. The detection of $\gamma$-ray emission may be indicative of an on-axis viewing angle, but the low $\gamma$-ray fluence \citep{GBMdetection} implies an isotropic-equivalent kinetic energy of only $E_{\rm K,iso}\sim 5\times 10^{47}$~erg (assuming an efficiency of $\eta_\gamma=0.1$), orders of magnitude lower than the energies inferred for cosmological SGRBs \citep{ber14,fbm+15}. From our radio observations {  at $2.4$~d} we rule out an on-axis afterglow with canonical parameters inferred from SGRBs of $E_{\rm K,iso}\sim 2\times 10^{51}$~erg and $n\sim 10^{-2}$~cm$^{-3}$ \citep{fbm+15}, for a wide range of values of $\epsilon_B$ (assuming $\epsilon_e=0.1$); see Figure \ref{fig:onaxis}. 

While all on-axis jets with $E_{\rm K,iso} \gtrsim10^{48}$ erg are ruled out, we find that for $E_{\rm K,iso}\sim 5\times 10^{47}$ erg our radio and X-ray upper limits can be accommodated for densities of $n\lesssim 1$ cm$^{-3}$ ($\epsilon_B=10^{-4}-10^{-2}$); see Figure \ref{fig:onaxis}. However, these low-density models predict fading emission and are therefore inconsistent with the radio detection at $\sim 16-40$ d and with the rising X-ray flux observed between 2.4 d and 15.4 d \citep{DECamPaper5}. We therefore conclude that an on-axis relativistic jet cannot explain our radio and X-ray detections. 


\subsection{Off-axis Afterglow Models}

\begin{figure*} 
\centering
\includegraphics[width=0.48\textwidth]{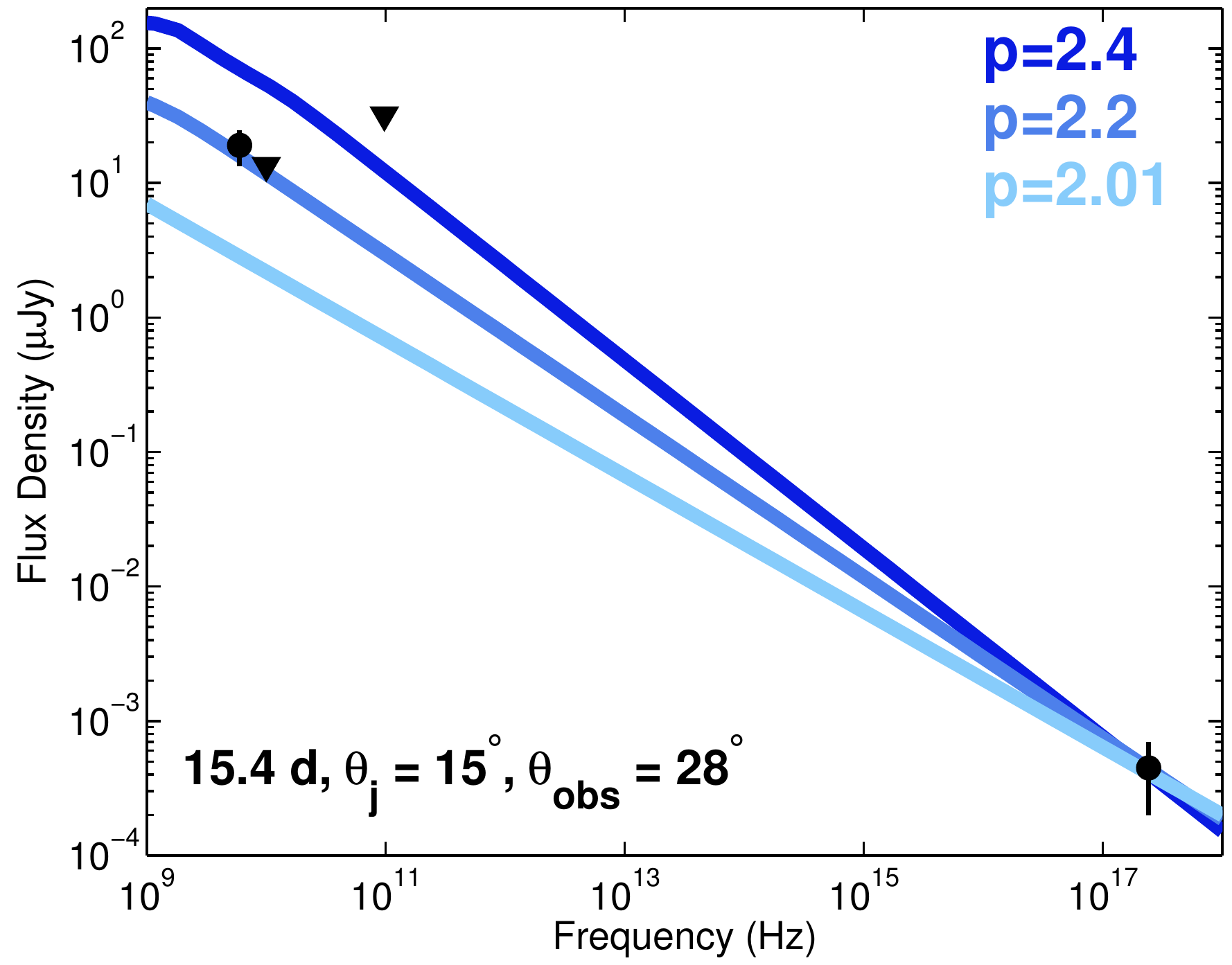}\includegraphics[width=0.48\textwidth]{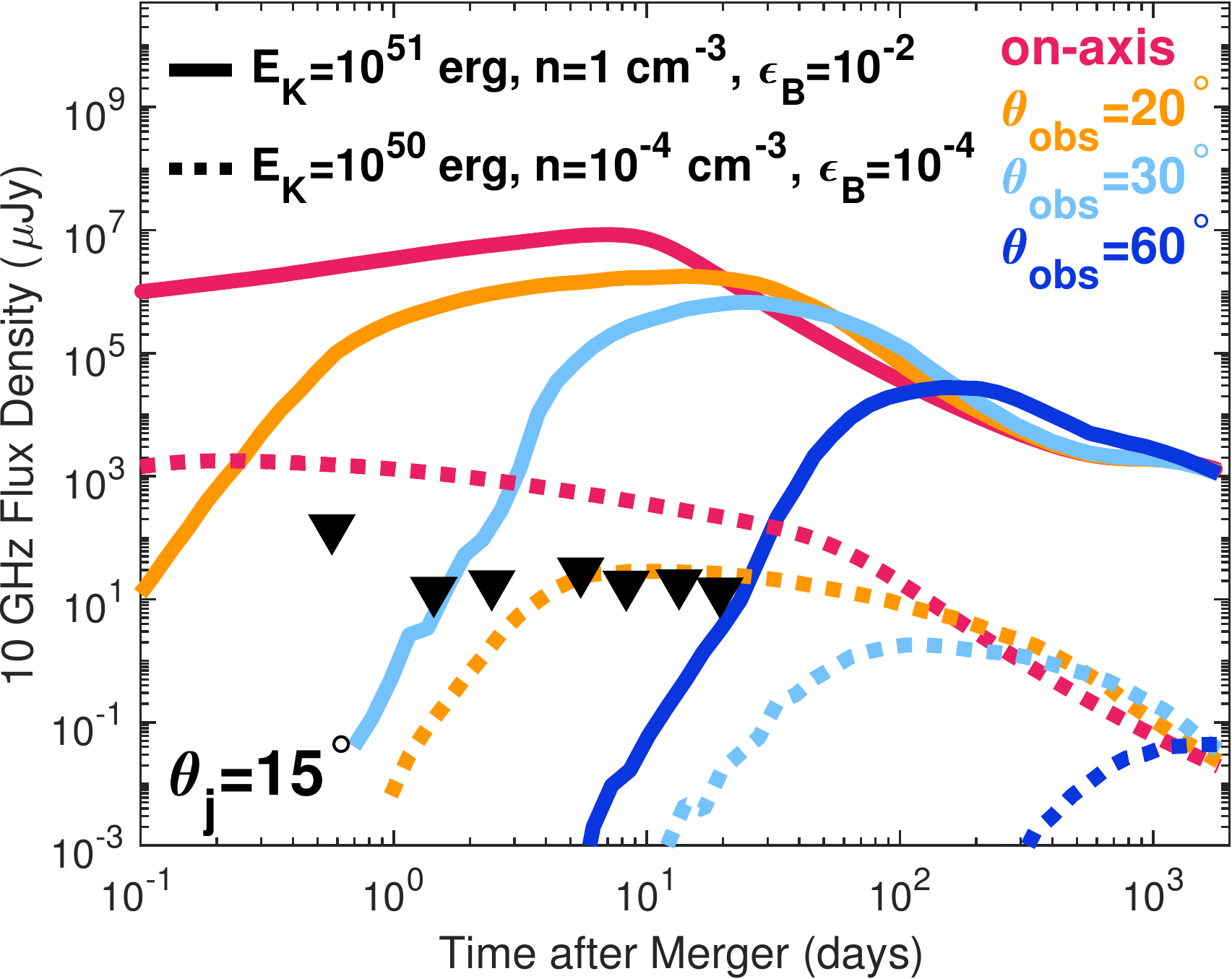}
\includegraphics[width=0.48\textwidth]{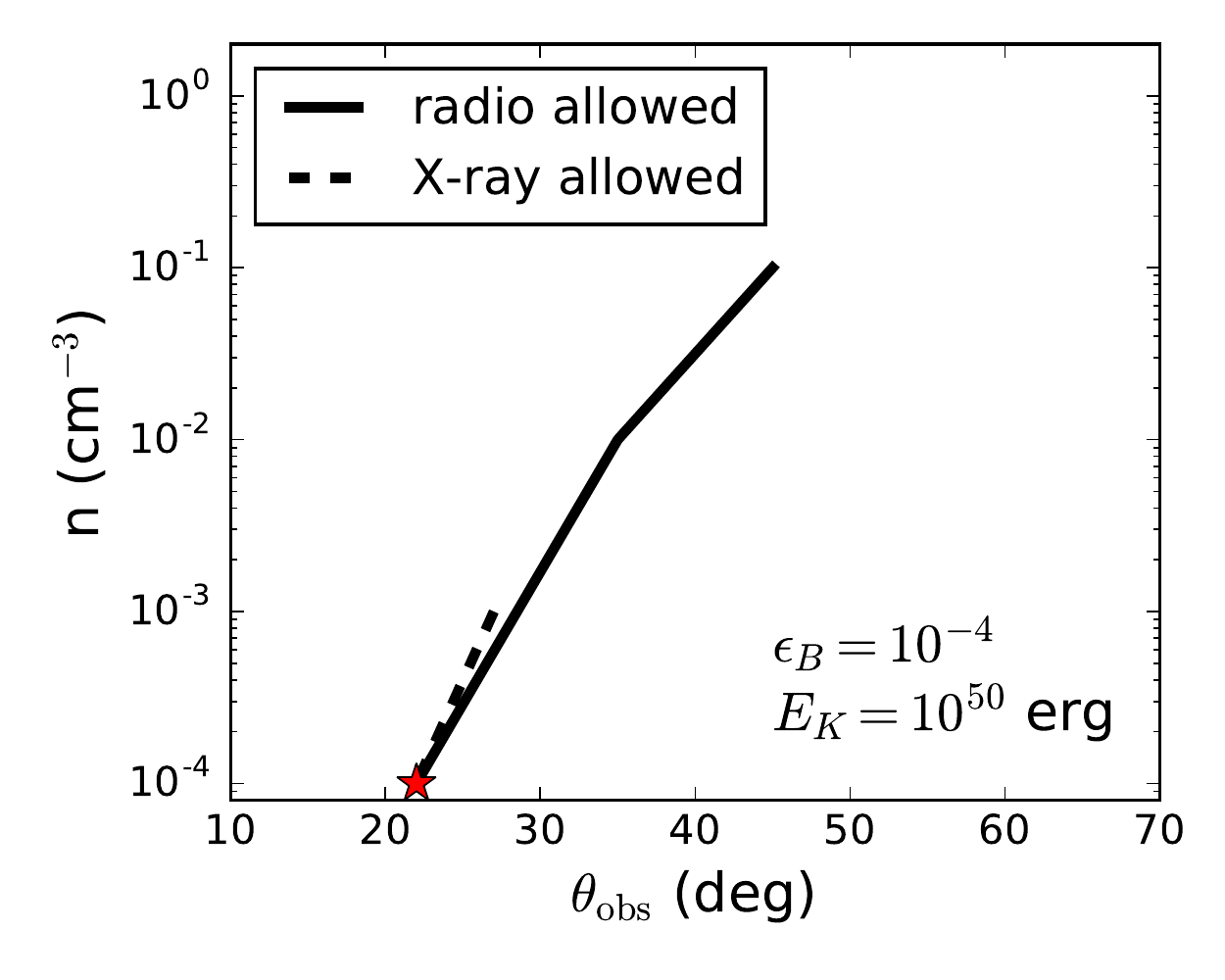}\includegraphics[width=0.48\textwidth]{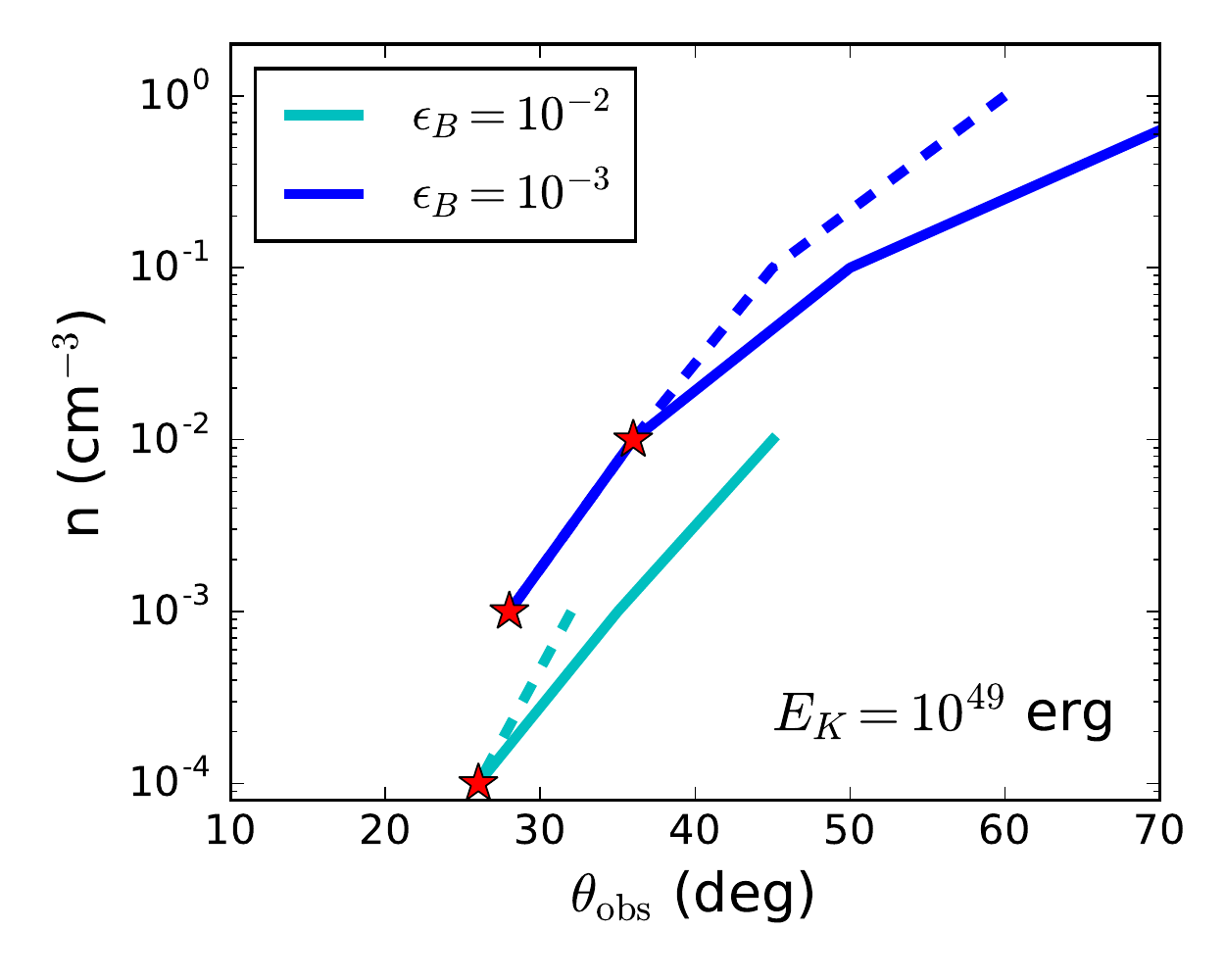}
\caption{Constraints on model parameters from the radio and X-ray data, based on the simulation set described in \cite{DECamPaper5}. Upper left: Three model SEDs of an off-axis $15^{\circ}$ jet at 15.4 d that fit the X-ray emission with different values of $p$. We find that models with $p>2.2$ that match the X-ray emission overpredict the radio emission, ruling out the median value for cosmological SGRBs, $p=2.4$. A model with $p=2.2$ matches our radio detection at 6 GHz. Upper right: Select model radio light curves explored in our simulations. Models with small $\theta_{\rm obs}$ are ruled out by our 10 GHz upper limits, while models with large $\theta_{\rm obs}$ are ruled out by our 6 GHz detections at 19 d and 39 d. Models with large $E_K$ are ruled out for all $\theta_{\rm obs}$ because they predict that the radio and X-ray flux densities increase faster than observed between $\sim10-40$ d. Bottom:  The regions of parameter space allowed by our radio {  and X-ray} observations for typical SGRB jet kinetic energies (assuming $p=2.2$ and $\epsilon_e=0.1$). The solid lines indicate the values of $n$ and $\theta_{\rm obs}$ allowed by the radio observations for fixed $\epsilon_B$ and $E_K$. Dashed lines of the same color indicate the corresponding values allowed by the X-ray observations. Red stars mark simulations that are consistent with both the X-rays and radio observations.}
\label{fig:offaxis}
\end{figure*}

We next explore models in which the radio emission originates from the afterglow of a relativistic SGRB jet viewed off-axis.  To constrain the value of $\theta_{\rm obs}$, we use the afterglow modeling code {\tt BOXFIT} (v2; \citealt{vzm+10,vm12}) for a wide range of kinetic energies, densities, jet opening angles, observer orientations, and $\epsilon_B$, as described in \cite{DECamPaper5}; we fix $\epsilon_e=0.1$.

We first consider $p=2.4$ and $\epsilon_B=0.01$, the median values for SGRBs \citep{fbm+15}. We find that models with $p=2.4$ which match the X-ray light curve \citep{DECamPaper5} consistently over-predict the radio emission at a comparable epoch (Figure \ref{fig:offaxis}; {  top} left panel). On the other hand, assuming a value of $p=2.2$ matches both the X-ray detection at 15 days and our observed weak radio emission at 6 GHz at 19 days, and is also consistent with our upper limits at 10 GHz and 97.5 GHz. All simulations discussed for the remainder of this paper assume $p=2.2$, but we explore wide ranges of $E_K$, $n$, $\epsilon_B$, and $\theta_{\rm obs}$ (Figure \ref{fig:offaxis}; top right panel).

When considered in isolation, the radio observations exhibit a strong degeneracy between $n$ and $\theta_{\rm obs}$, while varying $E_K$ and $\epsilon_B$ also causes shifts in the allowed parameter space (Figure \ref{fig:offaxis}; {  bottom panels}). Observer viewing angles $\theta_{\rm obs}\lesssim 20^{\circ}$ are ruled out, even for a low value of $\epsilon_B=10^{-4}$, while the largest viewing angles require densities $n\gtrsim 1$~cm$^{-3}$. Using the constraints from the X-ray observations shifts the allowed parameter space to low densities $n\sim 10^{-4}-10^{-2}$~cm$^{-3}$ and tightens the viewing angle constraint to $20^{\circ}\lesssim\theta_{\rm obs}\lesssim40^{\circ}$. Models with $E_K > 10^{50}$ erg and $\epsilon_B > 0.01$ are entirely ruled out.
This is consistent with our modeling of the optical and NIR emission \citep{DECamPaper2,DECamPaper3,DECamPaper4}, which suggests that $\theta_{\rm obs}\lesssim 45^{\circ}$ based on the presence of blue kilonova emission. The inferred values of $E_K$, $n$, and $\epsilon_B$ are well within the ranges of the observed populations of SGRBs \citep{fbm+15,DECamPaper7}. In particular, the inferred low density is consistent with GW170817's origin in an elliptical host galaxy\footnote{{  From surface brightness profile fitting, \citet{DECamPaper8} demonstrate that NGC\,4993 has an elliptical morphology.}}, as the expected ISM densities in elliptical galaxies are low \citep{fbp+06}.

\begin{figure*} 
\centering
\includegraphics[width=0.48\textwidth]{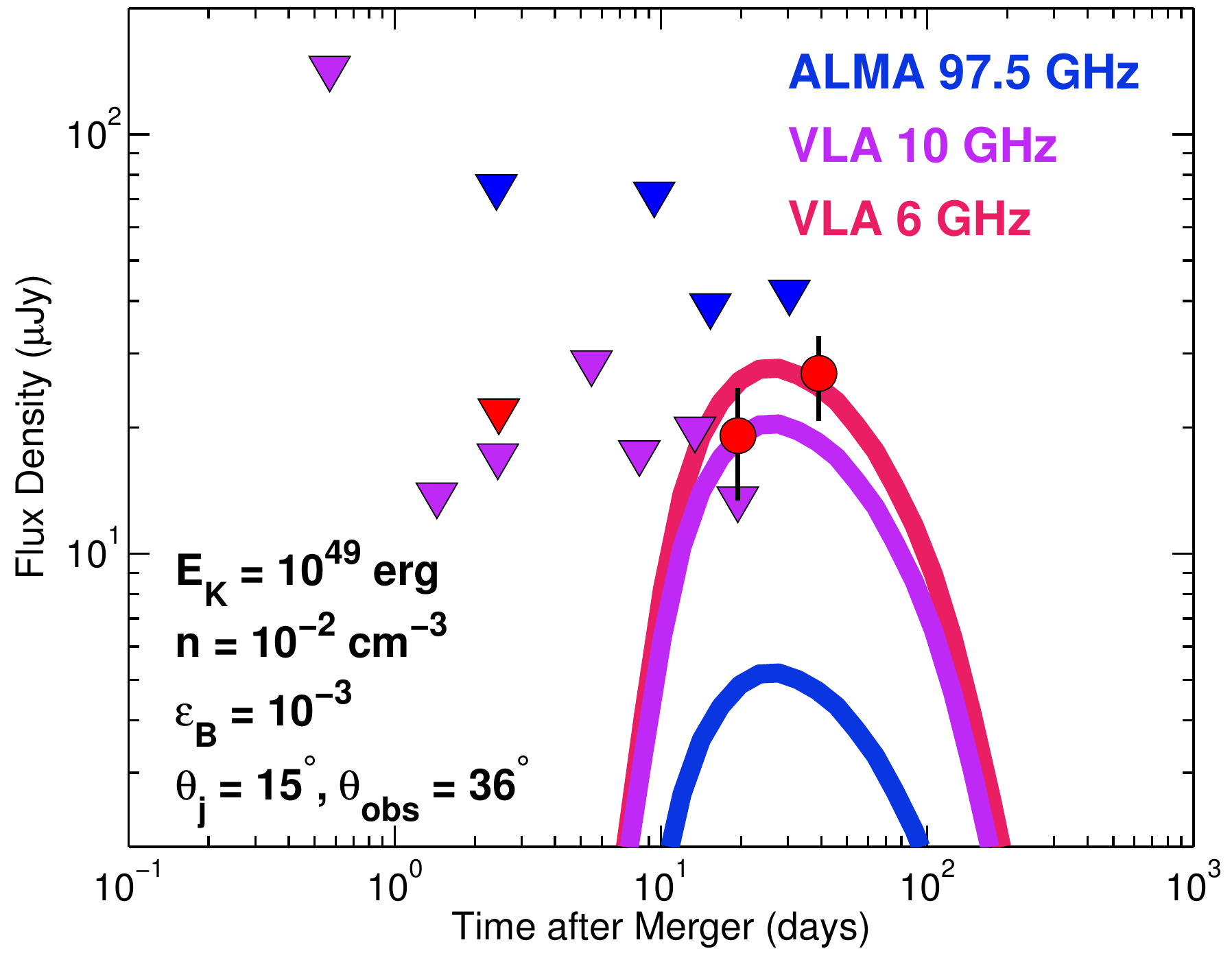}\includegraphics[width=0.48\textwidth]{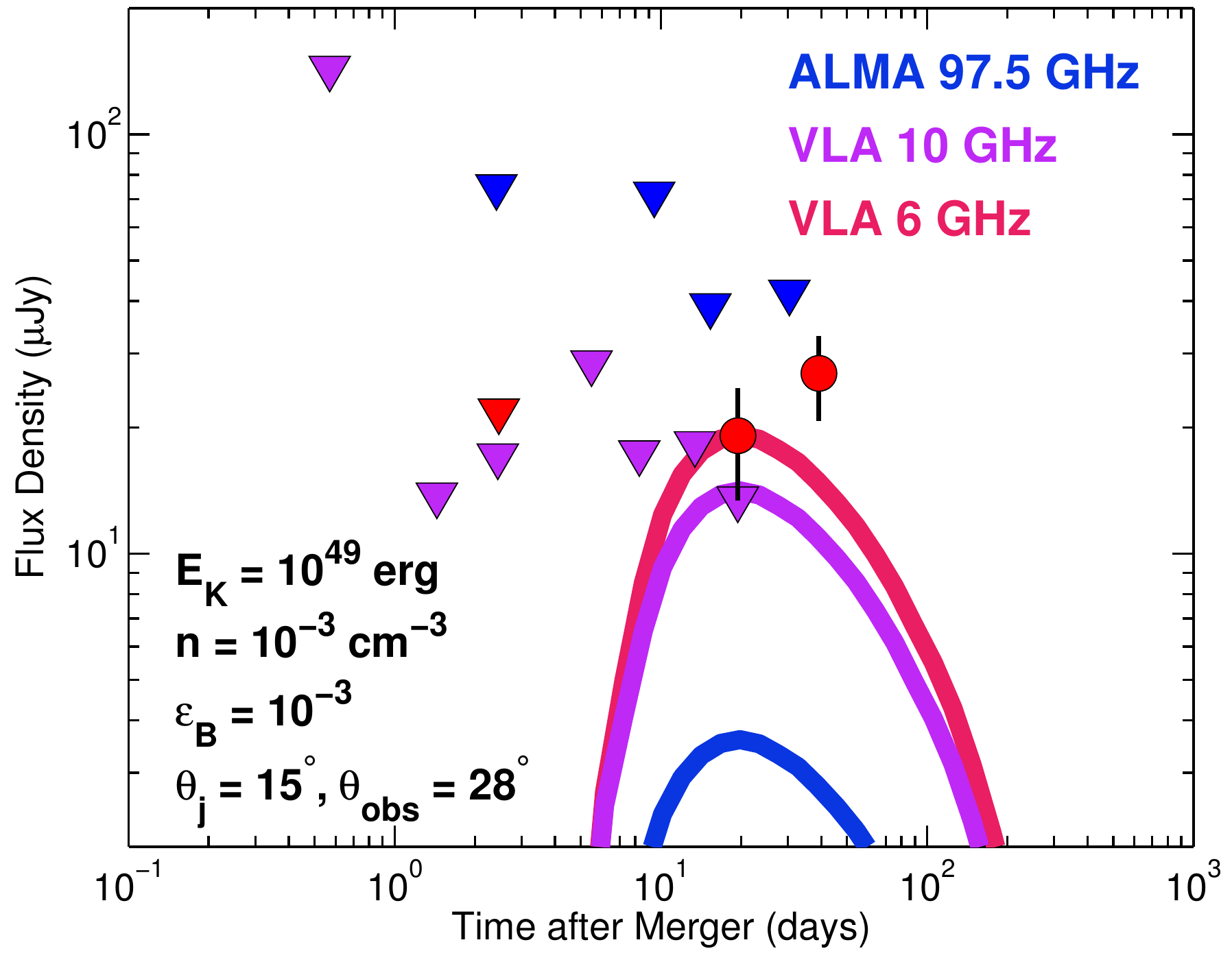}
\includegraphics[width=0.48\textwidth]{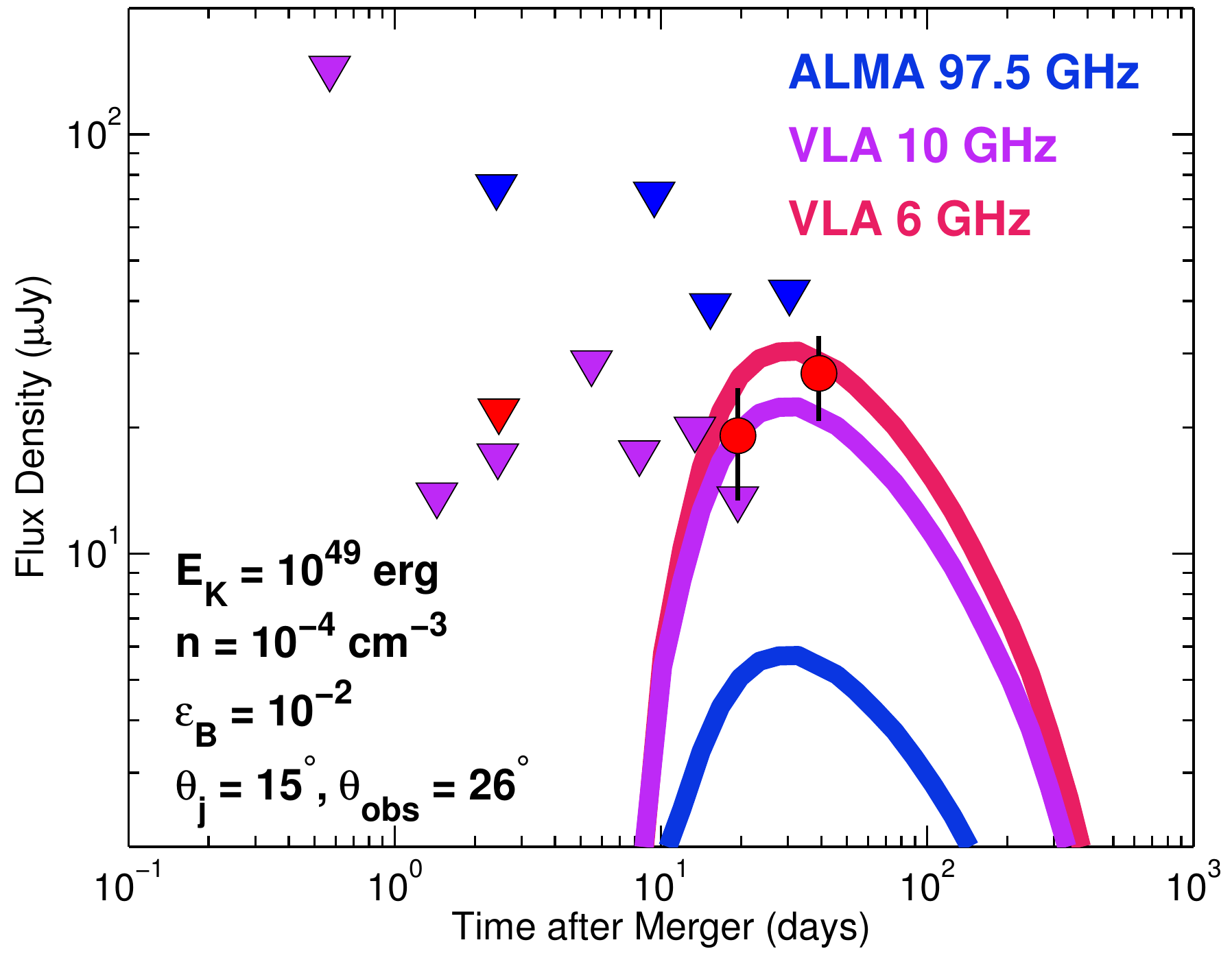}\includegraphics[width=0.48\textwidth]{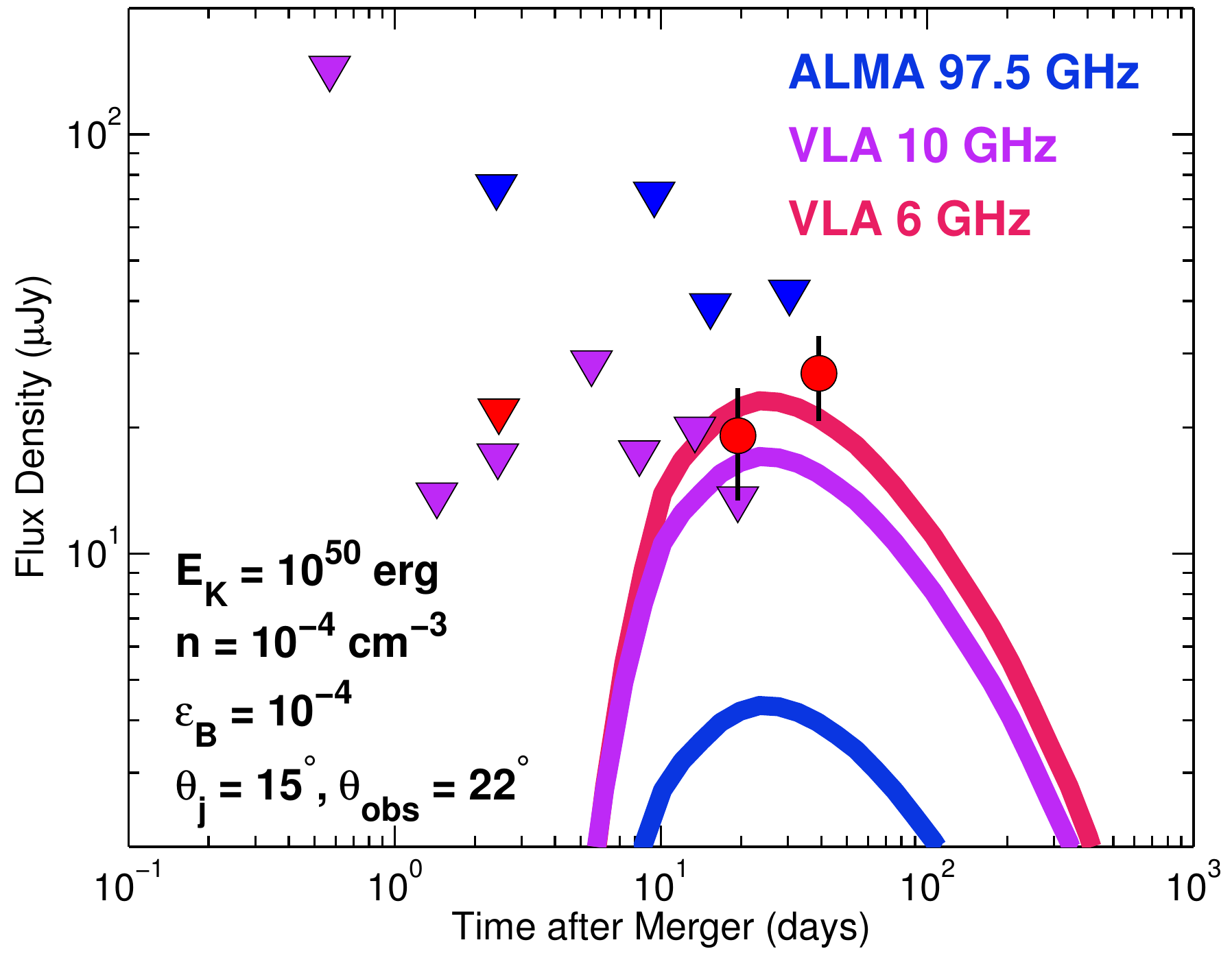}
\caption{Simulated radio light curves for the four models also presented in \citep{DECamPaper5}, shown with all of our radio upper limits (triangles; $3\sigma$) and detections (circles). The emission peaks on a timescale of $\sim15-30$ d, but should remain detectable at 6 GHz for {  weeks to} months. We note that the observations at 19.2 d were taken under poor weather conditions, which can lead to flux decorrelation at high frequencies of $\gtrsim10$ GHz. Our final 10 GHz upper limit may therefore underestimate the true flux density at this epoch.}
\label{fig:lc}
\end{figure*}

We find that all of the afterglow models that satisfy the current radio data peak on a timescale similar to the X-ray peak, $\sim 15-30$ d (Figure \ref{fig:lc}). As the peak is fairly broad, we predict that the emission should remain detectable with the VLA for weeks to months. As GW170817 is currently too close to the Sun to be observable by X-ray and optical facilities, radio observations will remain the only way to monitor the transient emission during this time. 
Continued radio monitoring of GW170817 will help us further narrow down this parameter space, allowing for tighter constraints on the burst energy and circumbinary density.

\section{Predictions for future radio emission from the kilonova ejecta}

In addition to the relativistic jet, BNS mergers are also expected to generate non-relativistic ejecta, which will produce synchrotron emission at radio wavelengths once it decelerates \citep{np11}. This is the same ejecta that initially generates the kilonova emission detected in the UV/optical/NIR bands. Compared to the relativistic jet, this ejecta component will decelerate on a significantly longer timescale due to its larger mass, $\approx 0.01-0.1$ M$_{\odot}$ \citep{mb14,hp15}. The radio emission from the kilonova ejecta is therefore expected to peak on timescales of months to years \citep{np11,mb12,mb14,hp15}. Searches for this component following a subset of cosmological SGRBs have all yielded deep non-detections, placing constraints on the kinetic energy injected of $\gtrsim 10^{51}$~erg in these events \citep{mb14,hhp+16,fmb+16}.

\begin{figure} 
\centerline{\includegraphics[width=3.5in]{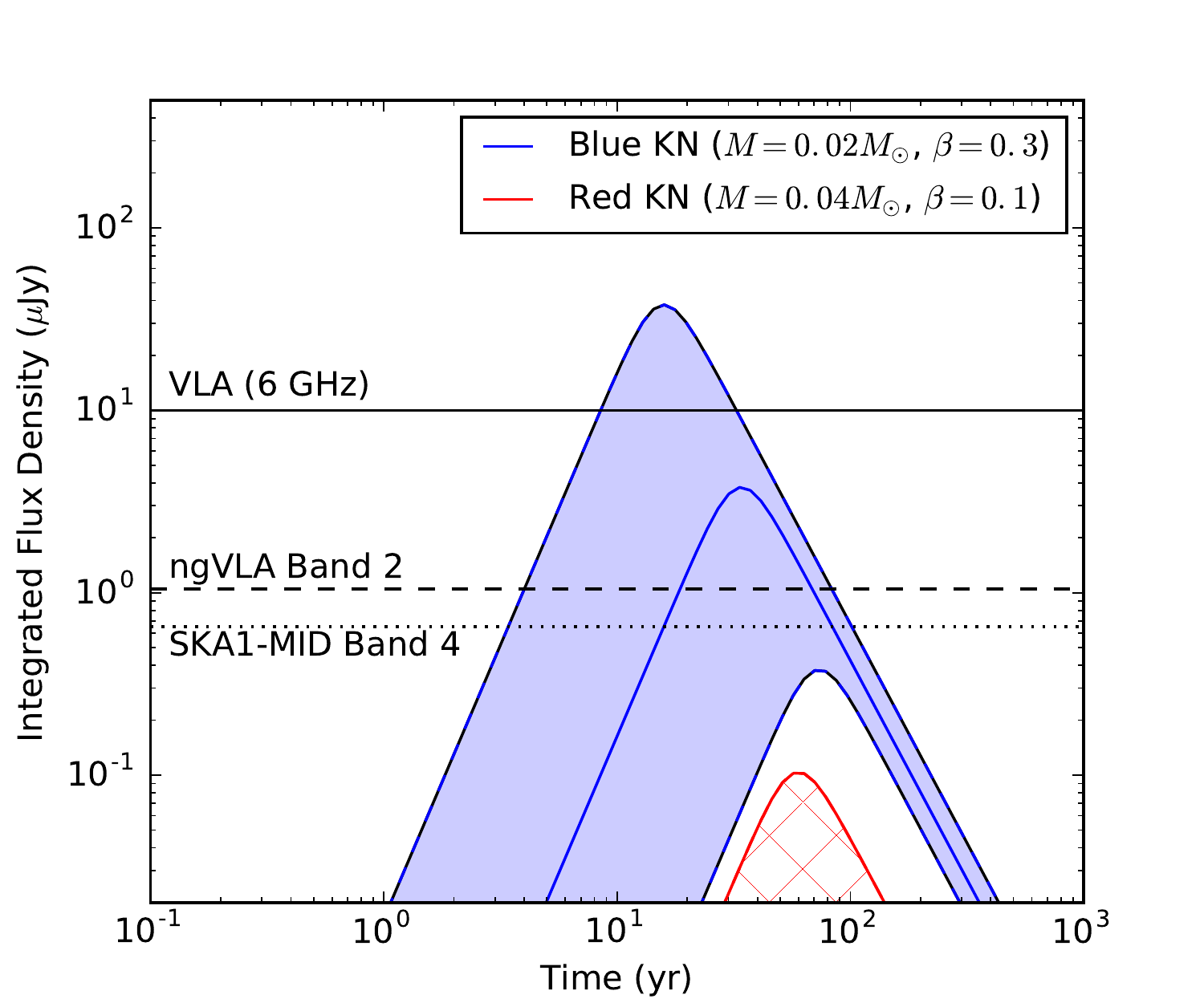}}
\caption{Radio emission predicted from decelerated kilonova ejecta for the two component model described in \cite{DECamPaper2} assuming the density range allowed by our VLA observations, $n=10^{-4}-10^{-2}$ cm$^{-3}$. The blue KN component (solid blue) is detectable by the VLA at its current sensitivity for favorable parameters and is easily detectable  for most of the allowed parameter range by the ngVLA and the SKA at design sensitivity, both of which are expected to be operational by the time the emission peaks. The red KN component (crosshatched red) takes longer to decelerate and is sub-dominant.}
\label{fig:kn}
\end{figure}

For the first time, we can make specific predictions for the kilonova radio emission using the parameters inferred from modeling of the UV/optical/NIR emission \citep{DECamPaper2,DECamPaper3,DECamPaper4}. The kilonova emission requires two components: a ``blue'' component with $M_{\rm ej}\approx 0.02$ M$_{\odot}$ and $v_{\rm ej}\approx 0.3c$, and a ``red'' component with $M_{\rm ej}\approx 0.04$ M$_{\odot}$ and $v_{\rm ej}\approx 0.1c$ \citep{DECamPaper2,DECamPaper3,DECamPaper4}. The predicted radio emission from each component is shown in Figure~\ref{fig:kn} for a fiducial density of $n=1\times10^{-3}$ cm$^{-3}$ (solid lines). The shaded bands indicate the full range of possible densities preferred by our modeling of the radio and X-ray counterparts to GW170817. 

We predict that the blue kilonova component will dominate the radio emission at all times and will be detectable with the VLA at its current sensitivity as early as $\sim 5$ yr post-merger for $n=10^{-2}$ {  cm$^{-3}$}.  This component dominates because of its larger kinetic energy and earlier deceleration time. {  For densities $n\lesssim3\times10^{-3}$ cm$^{-3}$ the blue kilonova will not be detectable with the current VLA, but} the next generation of sensitive radio telescopes, including ngVLA \citep{ngVLA} and SKA1-MID \citep{SKA} will be able to detect emission from this component for decades. Emission from the red kilonova component remains sub-dominant at all times. We note that radio emission from the KN ejecta-ISM interaction could begin even earlier than we have predicted if the ejecta contains a moderate tail of even faster expanding matter with velocity $\gtrsim0.3 c$, to which optical KN observations of GW170817 are not sensitive (since its optical/UV emission would have peaked on earlier timescales of a few hours; e.g. \citealt{Metzger+15, np16}). 

\section{Conclusions}

We presented extensive radio follow-up observations of GW170817 at centimeter and millimeter wavelengths, including the earliest observations taken in these bands. Our observations rule out a typical SGRB on-axis jet ($E_{\rm K,iso}\gtrsim10^{48}$ erg). Instead, we find that our radio observations, together with the X-ray light curve \citep{DECamPaper5}, can be jointly explained as the afterglow from an off-axis relativistic jet with an energy of $10^{49}-10^{50}$ erg expanding into a low-density medium of $\sim 10^{-4}-10^{-2}$ cm$^{-3}$, at an inferred $\theta_{\rm obs}\approx 20-40^{\circ}$. Under this interpretation, {  GW170817} would be the first detection of an off-axis afterglow from a SGRB, and would also be the first direct observational evidence for the launching of relativistic jets in BNS mergers. As the early optical emission is dominated by the kilonova ejecta, radio and X-ray observations will continue to be the best way to probe relativistic outflows in BNS mergers discovered by LIGO/Virgo, the majority of which will be off-axis (e.g., \citealt{mb12}).

We also use the kilonova ejecta properties inferred from our UV/optical/NIR data and modeling to place the first observationally-motivated constraints on the predicted radio emission from the non-relativistic ejecta. Detection of this component allows an independent measurement of the ejecta properties and the circumbinary density, but is more challenging than detection of the afterglow due to its longevity.  For GW\,170817 we predict emission from this component on a timescale of at least a few years post merger.  The next generation of radio telescopes will come online by the time the emission from GW170817 and future LIGO/Virgo BNS merger events reach their peak. In the upcoming era of high-sensitivity all sky radio surveys, radio emission from BNS mergers will become a powerful piece of the EM toolkit in the new field of multi-messenger GW-EM astronomy.

\acknowledgments
The Berger Time-Domain Group at Harvard is supported in part by the NSF through grants AST-1411763 and AST-1714498, and by NASA through grants NNX15AE50G and NNX16AC22G.
 WF acknowledges support from Program number HST-HF2-51390.001-A, provided by NASA through a grant from the Space Telescope Science Institute, which is operated by the Association of Universities for Research in Astronomy, Incorporated, under NASA contract NAS5-26555. CG acknowledges University of Ferrara for use of the local HPC facility co-funded by the ``Large-Scale Facilities 2010'' project (grant 7746/2011). Development of the Boxfit code was supported in part by NASA through grant NNX10AF62G issued through the Astrophysics Theory Program and by the NSF through grant AST-1009863. BDM is supported in part by NASA ATP grant NNX16AB30G. Simulations for BOXFITv2 have been carried out in part on the computing facilities of the Computational Center for Particle and Astrophysics of the research cooperation ``Excellence Cluster Universe'' in Garching, Germany. This paper makes use of the following ALMA data: ADS/JAO.ALMA\#2016.A.00043.T and ADS/JAO.ALMA\#2016.A.00046.T. ALMA is a partnership of ESO (representing its member states), NSF (USA) and NINS (Japan), together with NRC (Canada), MOST and ASIAA (Taiwan), and KASI (Republic of Korea), in cooperation with the Republic of Chile. The Joint ALMA Observatory is operated by ESO, AUI/NRAO and NAOJ. The National Radio Astronomy Observatory is a facility of the National Science Foundation operated under cooperative agreement by Associated Universities, Inc.

\software{CASA, Numpy, pwkit}



\end{document}